# DOCUMENT SUMMARIZATION USING POSITIVE POINTWISE MUTUAL INFORMATION

Aji S and Ramachandra Kaimal

Department of Computer Science, University of Kerala, Kerala, India
`aji_12345@yahoo.com`

School of Engineering, Amrita Vishwa Vidyapeetham Amritapuri Campus, Kollam, India, `mrkaimal@yahoo.com`

## ABSTRACT

*The degree of success in document summarization processes depends on the performance of the method used in identifying significant sentences in the documents. The collection of unique words characterizes the major signature of the document, and forms the basis for Term-Sentence-Matrix (TSM). The Positive Pointwise Mutual Information, which works well for measuring semantic similarity in the Term-Sentence-Matrix, is used in our method to assign weights for each entry in the Term-Sentence-Matrix. The Sentence-Rank-Matrix generated from this weighted TSM, is then used to extract a summary from the document. Our experiments show that such a method would outperform most of the existing methods in producing summaries from large documents.*

## 1. INTRODUCTION

The escalation of the computer networks and easy access methods to information has led to increasing amount of storage of information, mostly textual. According to the latest report from IDC [1], the world's information is doubling every two years. In 2011, the information created around the world was more than 1.8 zettabytes. By 2020 the world will generate 50 times the amount of information and 75 times the number of "information containers" while IT staff to manage it will grow less than 1.5 times. The report also points out the necessity of new "information taming" technologies for information processing and storage.

To speedup the accessing, the flow of information needs to be filtered and stored systematically. For example, the working of Information Retrieval Systems (IRS) can be made effective by summarizing the entire collection of documents. Automatic text summarization can help by providing condensed versions of text documents. Expected summarization holds a list of applications like information extraction, document retrieval [2], evaluation of answer books [3], etc.

Since the first study on text extraction by Luhn appeared, the text summarization process has attracted lot of research activities [14,16,17]. Depending on the purpose and intended users, a summary can be generic or user-focused [4]. A generic summary covers all major themes or aspects of the original document to serve a broad readership community rather than a particular group. A user-focused (or topic-focused, query oriented) summary favors specific themes.

Summarization processes are traditionally confined to ad-hoc and simple techniques, without any symbolic or linguistic processing, and this limits the quality of summary that can be produced. Semantic similarity is a concept whereby a set of words within identified unique words are assigned a metric based on the worthiness/ correctness of their meaning or semantic content. In this paper we suggest a method based on Positive Pointwise Mutual Information (PPMI) [5] an extension of Pointwise Mutual Information PMI[6] which gives more importance to measure the semantic similarity between the words in a document for document summarization.

## 2. METHOD

In linguistics, morphology [7] deals with the arrangement and relationships between the words in a document. In any type of text processing application, the first step will be morphological analysis. Tokenization, stop words elimination [8] and stemming [9] are the sub tasks that are followed in our method.

### 2.1 Tokenization and stop words elimination

Even though characters are the smallest unit, words are considered as the useful and informative building blocks of a document for processing. As depicted in the figure 1, the sentences in the document are separated and will be treated as the samples $S_i, i = 1,...n$ for the experiment. Words in $S_i$ are separated in the next step and the punctuation marks and other irrelevant notations will be removed from those words.

Stop words are very commonly used words like 'the', 'of', 'and', 'a', 'in', 'to', 'is', 'for', 'with', etc that do not contribute anything to the informational content of a document and hence it can be removed. These stop words have much meaning in natural language processing techniques that evaluate grammatical structures, but they have less importance in statistical analysis.

### 2.2 Stemming

Generally the morphological variants of words separated from a document have analogous semantic understandings and can be considered as equivalent in IR system. A couple of algorithms [Lovins Stemming, Porter Stemming] for stemming [10,11] have been developed to reduce a word to its stem or root. After the stemming process, the terms of a document are the stems rather than the original words. Stemming algorithms not only reduce a word into stem, but also reduce the size of the list of words that has to be considered for analysis.

We are following the Porter Stemming [11] method, which is a rule based algorithm that works with both suffixes and prefixes. The algorithm defines five successive steps each consisting of a set of rules for transformation.

Here a word is represented as combination of consonants and vowels in the form

$$[C]VCVC......[V] \tag{1}$$

where the sequence bracket denotes arbitrary presence of their content and this can be written as

$$[C](VC)_m[V] \tag{2}$$

,where m is the number of occurrence of VC.

The further processing of stripping is decided by the rules applied in various steps in the algorithm.

At the end of stemming process, the unique words,

$$U_j, j = 1,....ps_i$$

where $ps_i$ is the number of unique words, will be separated from $S_i$. After processing each sentence, the collection of unique words in the entire document $T_i, i = 1,...t$, where t is the total number of unique words identified for the document is obtained.

## 2.3 Term-Sentence-Matrix

The occurrence of t words in the document is represented by a Term-Sentence-Matrix (TSM) of n columns and t rows, where t is the number of unique words and n is the number of sentences in the entire document. Each element $F_{ij}$ of the matrix is suitably measure the importance of term i with respect to the sentence and the entire document. Initially $F_{ij}$ is the frequency of that ith term in the jth sentence.

## 2.4 Weighting the Elements

TSM alone is not adequate for analyzing the feature of a document; terms that have a large frequency are not necessarily more imperative. A weight derived in respect of the local and document context can give more information than a frequency.

Mutual Information (MI)[12] of an entry measures the amount of information contributed by that entry in the entire document. Consider a pair of outcomes x and y, say the occurrence of words x and y, the MI is defined as:

$$MI(x, y) = \log \frac{p(x, y)}{p(x)p(y)} \quad (3)$$

$$= \log \frac{p(x/y)}{p(x)} \quad (4)$$

$$= \log \frac{p(y/x)}{p(y)} \quad (5)$$

The measure is symmetric and can be positive or negative values, but is zero if x and y are independent.

$$-\infty \leq MI(x, y) \leq \min[-\log p(x), -\log p(y)]$$

The value of MI maximizes when X and Y are perfectly associated. The negative MI shows that the co-occurrence is too small. The Positive PMI (PPMI) [12] is a modified version of PMI, in which all MI values that are less than zero are replaced with zero [13].

Consider the TSM, F, with t rows and n columns. The row vector $W_i$ corresponds to the ith word and the column vector $S_j$ corresponds to the jth sentence..

An element $F_{ij}$ gives the number of occurrence of i$^{th}$ word in the j$^{th}$ sentence. The row $f_{i:}$ corresponds to a word $w_i$ and the column $f_{:j}$ corresponds to a context $S_j$. The PPMI value of an element can be calculated as

$$pw = \frac{f_{ij}}{\sum_{i=1}^{t}\sum_{j=1}^{n} f_{ij}} \qquad (6)$$

$$pwi = \frac{\sum_{j=1}^{n} f_{ij}}{\sum_{i=1}^{t}\sum_{j=1}^{n} f_{ij}} \qquad (7)$$

$$ps = \frac{\sum_{i=1}^{t} f_{ij}}{\sum_{i=1}^{t}\sum_{j=1}^{n} f_{ij}} \qquad (8)$$

$$p_{ij} = \log\left(\frac{pw}{pwi \cdot ps}\right) \qquad (9)$$

$$ppmi_{ij} = \begin{cases} p_{ij} & if\ P_{ij} > 0 \\ 0 & otherwise \end{cases} \qquad (10)$$

where pw is the probability that the word $w_i$ occurs in the sentence j with respect to the entire document, $pw_i$ is the probability of word $w_i$ in the entire documents and ps is the probability of a sentence in the entire document. If $w_i$ and $s_j$ are statistically independent, then $pwi \cdot ps = pw$, and thus $ppmi_{ij}$ is zero (since log(1) = 0). The product $pwi.ps$ is what we would expect for pw if $w_i$ occurs in $s_j$ by pure random chance. If there is semantic relation between $w_i$ and $s_j$, then the pw should be larger than it would be if $w_i$ and $s_j$ were independent; hence pw > pwi.ps , and $ppmi_{ij}$ is positive; otherwise $ppmi_{ij}$ should have a value zero.

## 2.5 Ranking the sentence

The total significance of kth sentence, sk, can be calculated from the PPMI matrix as

$$s_k = \sum_{i=1}^{t} PPMI_{ik} \cdot ps_k \qquad (11)$$

,where $ps_k$ is the probability of kth sentence in context of document to be summarized.

The sentences in the entire documents are ranked according to the $s_k$. The sentences with required percentage weight is identified, and arranged in the order of as it in the original document.

## 3. EXPERIMENTAL RESULTS

A bunch of top hit articles in the online edition of Washington post are collected for the experiment. The articles contain an average of 850 words and 45 sentences. These articles are stored as plain text. The implementation strategy of our method is explained in the figure 1.

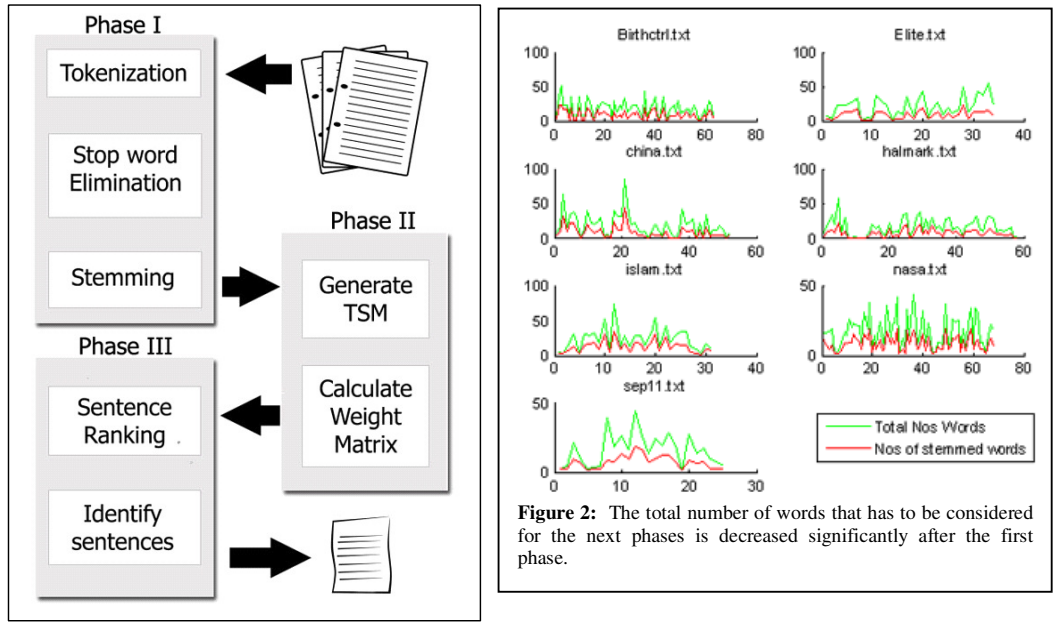

**Figure 2:** The total number of words that has to be considered for the next phases is decreased significantly after the first phase.

**Figure1:** There are three phases in the implementation; the document to be summarized is given to the tokenization process of first phase. The summary of the document will be outputted from the identify sentence process of the third phase.

Here we are considering seven documents for discussing the implementation details. The figure 2 explains the status of feature extracting process after the first phase.

Even if the total number of words before and after stemming has a well defined relation, the number of words after stemming has considerably decreased in each document.

An average of 50% of words is eliminated from each document in the first phase.

TABLE 1
DOCUMENT WISE WORD STATISTICS AFTER FIRST PHASE

| Document | Nos Sentences (n) | Total Nos of Words | Total Nos of words after elimination (t) | % words eliminated |
|---|---|---|---|---|
| Birthctrl.txt | 63 | 1121 | 572 | 49 |
| Elite.txt | 34 | 728 | 309 | 58 |
| china.txt | 52 | 994 | 468 | 53 |
| halmark.txt | 58 | 950 | 424 | 55 |
| islam.txt | 31 | 731 | 382 | 48 |
| nasa.txt | 68 | 1158 | 569 | 51 |
| sep11.txt | 25 | 408 | 180 | 56 |

The unique words identified in the first phase are used to create Term-Sentence-Matrix. Number of occurrence of ith word in jth sentence is the initial value of an entry, and naturally it will be 1 in most of the cases. The weight of each term in context of corresponding sentence and document are derived from the TSM using equations 6, 7, 8, 9 and 10.

The least significant elements in the TSM are eliminated while calculating the PPMI. The sentences are ranked according to the weight obtained in PPMI.

Weight of kth sentence, $s_k$ is calculated from the matrix PPMI using equation 11.

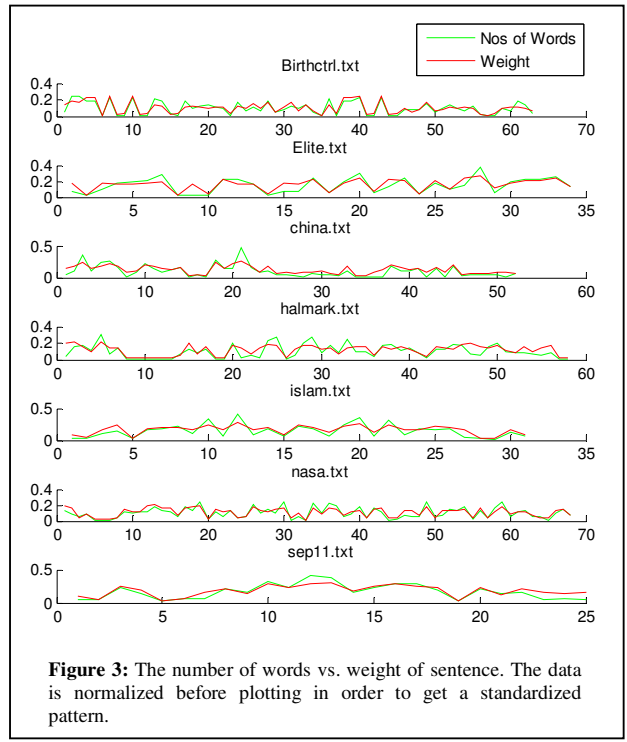

**Figure 3:** The number of words vs. weight of sentence. The data is normalized before plotting in order to get a standardized pattern.

Weight of a sentence is the direct measure of relevance of a sentence in a document. It is quite clear from the figure3 that in some cases, the weight of the sentence is not proportional to the number of words in it. For example, title is the first sentence in all documents used in the experiments, and the relevance of the words in the title is comparatively larger than other words in the remaining sentences.

Number of sentences required in the abstract is identified and extracts the sentences with higher importance from the original document. These sentences are arranged in order of original document to obtain the desired summary.

## 4. EVALUATION

There is no clear and standardized explanation for the question, what constitutes a good summary. Evaluation of summary is a major challenge in summarization systems. Researchers are working over the last decades to answer that complex question. Evaluation based on Latent Semantic Analysis[15] is new method in this area. This method evaluates the quality of summary through the content similarity between the document and its summary.

## 4.1 Measure of Main Topic

In addition to the existing PPMI matrix, we have constructed another matrix, SMI, for the summary from PPMI. SMI consist of t rows and l columns, where l is the number of sentences in the summary. The SVD method decomposes PPMI into three components as

$$PPMI = Ud\ Sd\ Vd^T \qquad (12)$$

and the SMI will be transformed as

$$MI = Us\ Ss\ Vs^T \qquad (13)$$

The first left singular vector of Ud is called the main topic[18] of the article. In this approach the main topic of both summary and document are calculated.

These vectors are the most significant features of the document and summary.

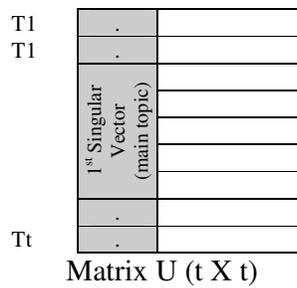

**Figure 4:** Main topic of a document

The classical cosine, $\cos\phi$, between the "main topic vector" of document and the "summary" reveals the degree of quality of the abstract.

$$\cos\phi = \sum_{i=1}^{t} ud_i \cdot us_i \qquad (14)$$

Where ud and us are the main topic of Ud and Us respectively. The following figure shows the final result of evaluation.

TABLE2:
$\cos\phi$ BETWEEN THE MAIN TOPIC OF DOCUMENT AND ITS SUMMARY

| Doc | Abstract in % | | | | |
|---|---|---|---|---|---|
| | 10 | 15 | 20 | 25 | 30 |
| D1 | 0.9994 | 0.9998 | 0.9999 | 1 | 1 |
| D2 | 0.9637 | 0.9981 | 0.9981 | 0.9983 | 1 |
| D3 | 0.9942 | 0.9972 | 0.9972 | 0.9998 | 1 |
| D4 | 0.9973 | 0.999 | 1 | 1 | 1 |
| D5 | 0.971 | 0.9716 | 0.9696 | 0.9772 | 0.9985 |
| D6 | 0.9971 | 0.9947 | 0.9998 | 0.9999 | 1 |
| D7 | 0.9422 | 0.9348 | 0.8318 | 0.9981 | 0.9981 |
| AVG | 0.9807 | 0.985 | 0.9709 | 0.9962 | 0.9995 |

International Journal of Computer Science & Information Technology (IJCSIT) Vol 4, No 2, April 2012

The result given in table2 says that, as a general trend the difference between the features of documents and its abstract reduces on increasing the size of the abstract.

The average value of the similarity, the overall degree of success of the method, measure $\cos\phi$ for the entire documents in the five test cases (% of abstract - 10 to 30) is 0.98646, which shows that the positive point wise mutual information technique gives a promising result in the connection with the main topic evaluation strategy.

## 5. CONCLUSION

The proposed summarization method contains three separate phases. The porter stemming algorithm in the morphological analysis phase has reduced the feature matrix considerably. The Positive Point Mutual Information technique is used to find out the weight of sentences in a document. It is shown here, that the Latent Semantic Analysis is a reliable summary evaluation mechanism. It is noted that summary of some document reaches its maximum result in the very initial stages of experiments. The overall average value of $\cos\phi$, the distance measure between the main topics of summary and document, reveals that the importance of Positive Point Mutual Information in text data analysis and especially in summarization process.